# Hyperbolic illusion of the Mid-Holocene turning point

Ron W Nielsen aka Jan Nurzynski[*]

Environmental Futures Centre, Gold Coast Campus, Griffith University, Qld, 4222, Australia

Using the growth of population in Australia in the past 10,000 years it is illustrated here how an illusion created by hyperbolic distributions may lead easily to incorrect conclusions. Contrary to the published claim, there was no change in the mechanism of growth of the ancient human population in Australia around 5000 years before present (BP). Data for the number of rock-shelter sites, interpreted as reflecting the growth of human population, are analysed. They are shown to be monotonically increasing with time without any sign of intensification in the past 10,000 years. The growth of human population in Australia is in excellent agreement with the similar pattern describing the growth of the world population.

Johnson and Brook (2011) analysed the time-dependent distribution of the number of rock-shelter sites in Australia, which they interpreted as representing the growth of ancient human population. The data are displayed in Fig. 1. The time arrow is from right to left. These data seem to suggest a slow growth before 5000 years ago and a faster growth after that year.

[*] r.nielsen@griffith.edu.au; ronwnielsen@gmail.com; http://home.iprimus.com.au/nielsens/ronnielsen.html

Suggested citation:

Nielsen, R. W. aka Nurzynski, J. (2013). *Hyperbolic illusion of the Mid-Holocene turning point*. http://arxiv.org/ftp/arxiv/papers/1308/1308.6210.pdf

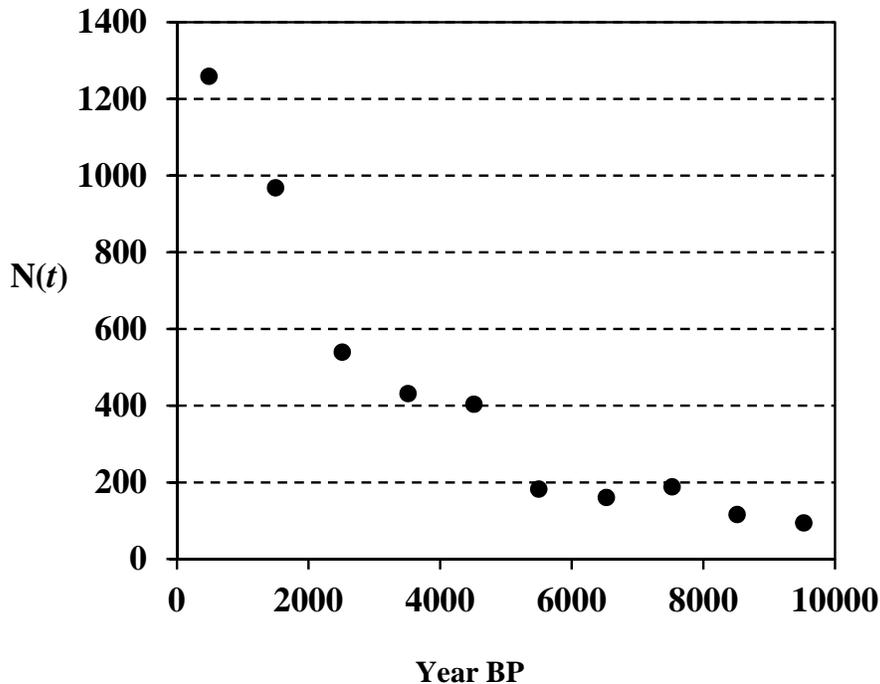

**Fig. 1.** Data for the relative number of rock-shelter sites (Johnson & Brook, 2011) representing the growth of human population in Australia, appear to show a dramatic change in the growth pattern (from slow to fast) around 5000 years before present (BP).

Johnson and Brook (2011) concluded that the growth of human population was "slow or negligible before 5000 years BP, and faster since then" (Johnson & Brook 2011, p. 1752). This observation led them inevitably to the question what might have triggered such a dramatic change in the growth pattern. "Whatever the trigger, our results provide new support for the view, advocated by some Australian archaeologists but contested by others, that something important happened to the human population of Australia during the Holocene, and that the Mid-Holocene in particular was a turning point in Australian prehistory" (Johnson & Brook 2011, p. 1753).

So now, the vital question is: Is their conclusion acceptable? Was there or was there not a significant alteration in the growth pattern of human population in Australia in the distant past? Was there really a turning point in the Australian prehistory? Did something important happen during the Holocene?

If there was a change, we have a research field wide open and we can look for answers? However, if the interpretation of data was in some way incorrect and if there was no change, we will have saved a great deal of time, effort and financial resources by not pursuing the suggested line of research. We can then divert our efforts into more promising channels.

The description published by Johnson & Brook (2011) sounded so familiar that it was hard to ignore it, because it closely reflected a common and easy-to-make mistake of seeing two different stages in certain types of monotonically increasing and perfectly undisturbed distributions. Their claim had to be investigated if not for any other reason than to see whether they have also made the same mistake. However, more importantly, it was to see whether there is a strong justification for continuing the line of research suggested by their conclusions.

The type of distributions analysed by Johnson & Brook (2011) may contain a genuine change in the pattern of growth but often they do not, and great care has to be taken to make sure that the interpretation of data is correct because the perceived two stages of growth might be just a strongly misleading illusion. In such cases, the transition between two alleged but non-existing stages is often described as an escape, sprint, sudden spurt, intensification, explosion, or by some other similar terms all emphasising a clear and dramatic change in the pattern of growth. It is amazing how many people fall into this trap (see for instance Clark 2003, 2005; Currais, Rivera & Rungo 2009; Galor 2005a, 2005b, 2007, 2008, 2012; Galor & Moav 2001; Galor & Mountford 2006; Galor & Weil 1999, 2000; Goodfriend & McDermott 1995; Hansen & Prescott 2002; Khan 2008; Klasen & Nestmann 2006; Kögel & Prskawetz 2001; Komlos 1989, 2000, 2003; Lagerlöf 2003a, 2003b, 2006, 2010; Leibenstein 1957;

Lucas 2002; Manfredi & Fanti 2003; Mataré 2009; McKeown 2009; McNeill 2000; Nelson 1956; Omran 1971, 1983, 1986, 1998, 2005; Robine 2001; Šimurina and Tica 2006; Smil 1999; Tamura 2002; Thomlinson 1965; van de Kaa 2008; Wang 2005, Warf 2010; Weisdorf 2004; Weiss 2007). The next and even more critical step is then to try to *explain* the mechanism of the two perceived stages of growth and of the associated transition by proposing distinctly different mechanisms for each of these imagined components. This step leads progressively further away from the correct understanding of the studied process because all efforts are now concentrated on the explanation of the non-existing phenomena. An increasing number of scholars are being involved. They do not question or dispute the existence of two stages of growth but only their explanations, proposing new mechanisms, hypotheses, theories and mathematical descriptions without realising that the apparent distinctly different two stages are just an illusion created by a certain types of monotonically increasing distributions.

Many ways can be used to cross-check our conclusions. One of the simplest ways is to look at the data from a new angle. For instance, when data vary over a large range of values, as in this case, it is helpful to present them using semi-logarithmic scales.

Data analysed by Johnson & Brook (2011) are presented in Fig. 2 using logarithmic scale for the vertical axis. We can see now clearly that, looking along the arrow of time, i.e. from right to left, they follow a monotonically increasing trajectory with no sign of any unusual acceleration or intensification. The two phases of growth, fast and slow, did not exist. There was no transition from a slow to a fast growth and there was nothing unusual in the growth pattern around 5000 years BP. Trying to explain the unusual change in the number of sites around that time or around any other time is totally irrelevant because there was no change.

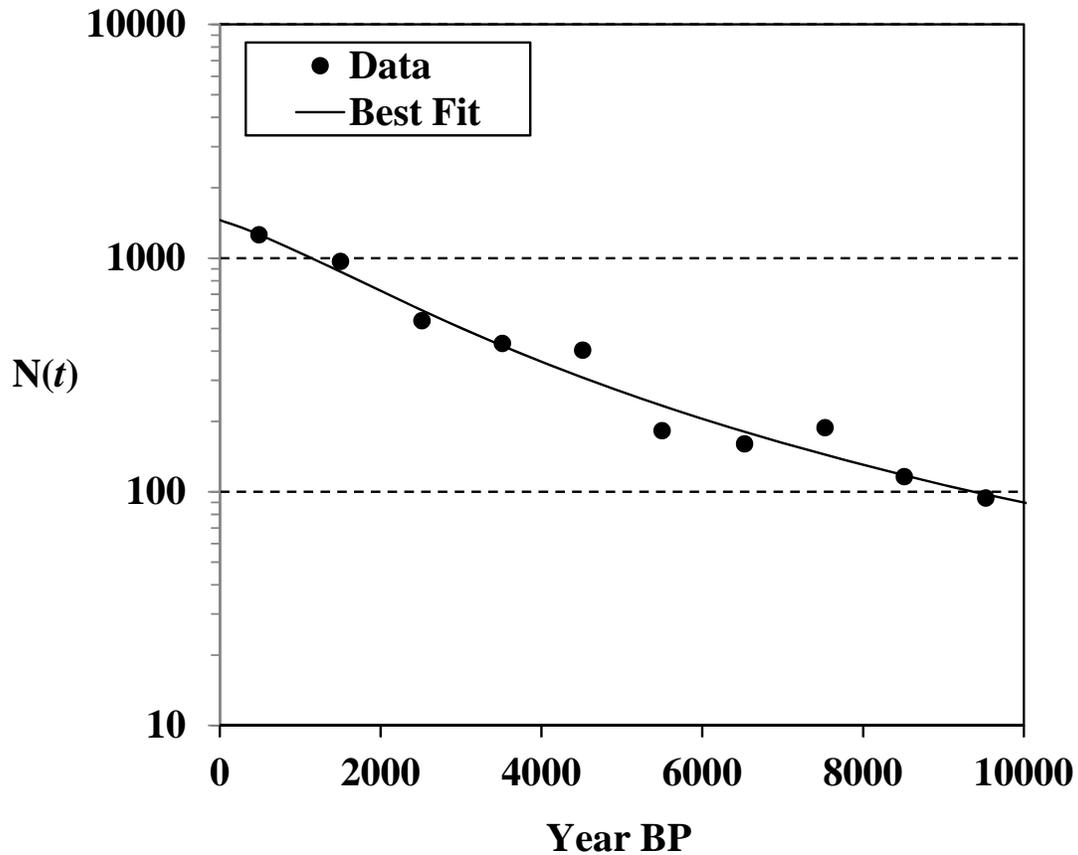

**Fig. 2.** The number of rock-shelter sites $N(t)$ shown in Fig. 1 is now plotted using semilogarithmic scales. The illusion of a sudden change around 5000 years BP has now disappeared. The best fit to the date is given by the second-order hyperbolic distribution. There was no change in the pattern of growth around 5000 years BP or at any other time.

Another useful way in trying to understand data is to plot and analyse their reciprocal values (Fig. 3). In this representation, an unusual acceleration or intensification would be shown as a clear *downward* change in the growth pattern. The reciprocal values of the data for the number of rock-shelter sites in Australia decrease monotonically with time without any signs of acceleration or intensification. There was no dramatic change and no unusual acceleration in the growth of the ancient population in Australia around 5000 years BP or at any other time.

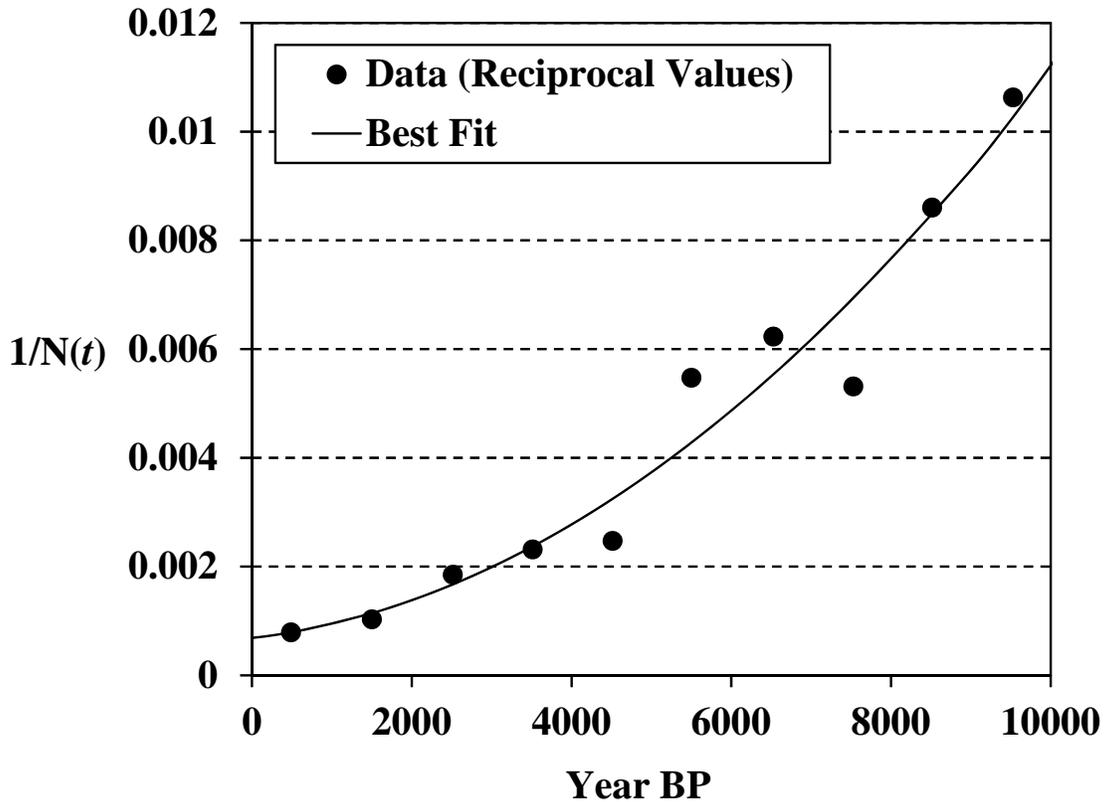

**Fig. 3.** The reciprocal values, $1/N(t)$, of the number of rock-shelter sites in Australia demonstrate also that there was no change in the pattern of growth around 5000 years BP or at any other time. Mathematical description of $N(t)$ is the same as in Fig. 2.

So now the puzzling conundrum, acknowledged by some Australian archaeologists (Lourandos 1997) but contested by others (Hiscock 2008) has been positively solved, and the approach is so simple, just a different way of plotting the same set of data. Nothing "important happened to the human population in Australia during the Holocene" (Johnson & Brook, p. 3753) and there was "no turning point in Australian prehistory" (Johnson & Brook, p. 3753), at least no turning point with respect to the number of rock-shelter sites and to the growth of the ancient human population. There was no trigger and no transition requiring explanation.

We can now go a step further and describe the data mathematically. It turns out that the best fit using the simplest mathematical description is given by the second order hyperbolic distribution

$$N(t) = \left(a_0 + a_1 t + a_2 t^2\right)^{-1} \tag{1}$$

where $t$ is the time in years BP, $N(t)$ is the number of rock-shelter sites interpreted as being proportional to the size of human population, $a_0 = 0.0006875$, $a_1 = 1.72 \times 10^{-7}$ and $a_2 = 8.7468 \times 10^{-11}$.

The growth of human population in Australia over the past 10,000 years was remarkably stable and was following closely a hyperbolic distribution, reflecting a similar pattern of growth as reported for the growth of the world population (Johansen & Sornette 2001; Karev 2005; Korotayev, 2005; Shklovskii 2002; von Foerster, Mora & Amiot 1960; von Hoerner 1975). Splitting this monotonically increasing distribution into two distinct components and trying to explain them by assuming different mechanisms of growth is not only unnecessary but also incorrect. There is nothing to explain about the change in the mechanism of growth because there was no change. However, the data suggest a remarkable and perhaps unexpected feature which could be further investigated. Why was the growth of the ancient human population in Australia so stable, so robust and so resilient to any variable forces over such a long time?